\documentclass[fleqn,usenatbib,useAMS]{mnras}

\usepackage{newtxtext,newtxmath}
\usepackage[T1]{fontenc}
\DeclareRobustCommand{\VAN}[3]{#2}
\let\VANthebibliography\thebibliography
\def\thebibliography{\DeclareRobustCommand{\VAN}[3]{##3}\VANthebibliography}

\usepackage{amsmath}
\usepackage{graphicx}
\def\ltsim{\lower.5ex\hbox{$\; \buildrel < \over \sim \;$}}

\def\bkt#1{\left(#1\right)}	
\def\bkts#1{\left[#1\right]}	
\def\D{\,\textrm{d}}
\def\diff#1#2{\frac{\textrm{d}{#1}}{\textrm{d}{#2}}}
\def\sub#1{_{\mbox{\scriptsize{#1}}}}
\def\super#1{^{\mbox{\scriptsize{#1}}}}

\title{The most massive Population III stars}

\author[T. Chantavat et al.]{
Teeraparb Chantavat,$^{1}$\thanks{E-mail:teeraparbc@nu.ac.th}
Siri Chongchitnan,$^{2}$
and Joseph Silk$^{3,4,5}$
\\
$^{1}$Institute for Fundamental Study, Naresuan University, Phitsanulok, 65000, Thailand\\
$^{2}$Warwick Mathematics Institute, University of Warwick, Zeeman Building, Coventry, CV4 7AL, United Kingdom\\
$^{3}$Institut d'Astrophysique de Paris, 98 bis Boulevard Arago, 75014, Paris, France \\
$^{4}$William H. Miller III Department of Physics and Astronomy,The Johns Hopkins University, Baltimore, MD 21218, USA\\
$^{5}$BIPAC, Department of Physics,University of Oxford, Keble Road, Oxford OX1 3RH, UK\\
}

\date{Accepted XXX. Received YYY; in original form ZZZ}

\pubyear{2023}

\begin{document}
\label{firstpage}
\pagerange{\pageref{firstpage}--\pageref{lastpage}}
\maketitle

\begin{abstract}
Recent data from the James Webb Space Telescope suggest that there are realistic prospects for detecting the earliest generation of stars at redshift $\sim20$.  These metal-poor, gaseous Population III (Pop III) stars are likely in the mass range $10-10^3\,{\rm M}_{\odot}$.  We develop a framework for calculating the abundances of Pop III stars as well as the distribution of the most massive Pop III stars based on an application of extreme-value statistics.  Our calculations use the star formation rate density from a recent simulation to calibrate the star-formation efficiency from which the Pop III stellar abundances are derived.  Our extreme-value modelling suggests that the most massive Pop III stars at redshifts $10<z<20$ are likely to be $\gtrsim10^3-10^4\,{\rm M}_\odot$.  Such extreme Pop III stars were sufficiently numerous to be the seeds of supermassive black holes at high redshifts and possibly source detectable gravitational waves.  We conclude  that the extreme-value formalism provides an effective way to constrain the stellar initial mass function.

\end{abstract}


\section{Introduction}
\label{sec:introduction}

The hypothetical first generation of stars, so-called  Population III (\textit{Pop III}), have long been  anticipated in the literature as massive, short-lived stars created in extremely metal-poor environments  \citep{Schwarzchild_Spitzer1953, Bond1981, Cayrel1986, Carr1994}.  Due to the lack of direct observational evidence, details of the physical properties of Pop III are not precisely known.  Many studies suggest that Pop III stars were formed within minihaloes of typical mass $\sim 10^6\,{\rm M}_\odot$ between redshift $z \sim 20 - 30$ and have a mass range between $10 - 10^3\,{\rm M}_\odot$ \citep{Haiman_ea1996, Tegmark_ea1997, Abel_ea2002, Bromm_ea2002, Yochida_ea2003, O'Shea_Norman2007, Susa_ea2014}.

Interest in Pop III stars has grown recently due to current and upcoming experiments that could potentially detect Pop III stars. These include the \textit{James Webb Space Telescope} (\textit{JWST}; \citet{Gardner_ea2006}), \textit{Euclid} \citep{Laureijs_ea2011, Marchetti_ea2017}  and the \textit{Roman Space Telescope} (\textit{RST}; \citet{Spergel_ea2015}).  Confirmed observations of Pop III stars would  solidify our understanding of stellar formation and evolution. However, the photometric signals from Pop III stars are expected to be very faint and would be extremely difficult to detect unless fortuitously enhanced by strong gravitational lensing \citep{Zackrisson_ea2012, Vikaeus_ea2022}.

Pop III stars are believed to end their lives in one of three channels: asymptotic giant branch stars, supernovae or black holes,  depending on their masses.  If the supernova progenitors  have mass in the range $\sim140-260\,{\rm M}_\odot$, they will ultimately form pair-instability supernovae (PISNe) that are unique to Pop III star evolution \citep{Moriya_ea2019}.  If the Pop III progenitors are sufficiently massive, their collapse will also emit highly energetic gamma-ray bursts  and after-glow components \citep{Bromm_Loeb2006, Kinugawa_ea2019}.  Another possibility is that massive Pop III stars evolve with accretion rates of $0.1-1\,{\rm M}_\odot$ yr$^{-1}$ until gravitational instability triggers their collapse to black holes \citep{Latif_ea2013, Inayoshi_ea2014, Umeda_ea2016, Becerra_ea2018, Haemmerle_ea2018}.
 
Given the potential of Pop III stars to give rise to early massive black holes, Pop III stars may help us understand a longstanding conundrum in astrophysics: the origin of quasars at very high redshift $z \gtrsim 6$ \citep{Fan_ea2001, Willott_ea2010, Mortlock_ea2011, Matsuoka_ea2019, Onoue_ea2019, Das_ea2021}. Such high-redshift quasars are associated with  supermassive black holes (SMBHs) with $M \gtrsim 10^9\,{\rm M}_\odot$ \citep{Volonteri2010, Inayoshi_ea2020}, which in turn could be seeded by  Pop III stars with mass $M \sim 10^3 - 10^5\,{\rm M}_\odot$ that  formed at redshift  $z \gtrsim 10$.  Such massive Pop III stars (which we  call \textit{extreme Pop III stars}) are certainly rare since most Pop III stars are expected to have mass $M \ltsim 10^2 \,{\rm M}_\odot$ and are difficult to grow into SMBHs via accretion processes and mergers \citep{Haiman_Loeb2001, Haiman2004, Volonteri2010}. 

However there are many uncertainties in the formation channels of Pop III stars, and they may cover a wide mass range \citep{Klessen_Glover2023}.  In rare cases, the most massive objects form by direct collapse and such extreme Pop III stars are  subject to general relativistic instabilities and can generate potentially detectable supernovae for precursors in a mass range around $3 \times 10^4\, {\rm M}_\odot$ \citep{Nagele_ea2022}.  Hence in our ensuing discussion, rather than attack the uncertain physics of Pop III star formation,  we will use a novel statistical approach to study the rarity of the most massive  Population III stellar objects based on empirical constraints.

To summarise the importance of studying extreme Pop III stars:
\begin{enumerate}
    \item From an observational point of view, the first Pop III stars to be directly detected are  likely to be amongst the most massive ones. The discovery of such objects will help us understand structure formation in the early universe and physics of the reionization epoch.
    
    \item Extreme Pop III stars can explain the origin of SMBHs at high redshifts. If observed, follow-up observations will give  us a  better understanding of the environment and the conditions for SMBH formation.
\end{enumerate}

In this work, we will demonstrate a formalism to calculate the mass distribution of the most massive Pop III stars based on  extreme-value statistics. Our technique involves a novel calculation of  \textit{star formation rate density} (hereafter SFRD) which we discuss below. 

There is no precise, universally agreed definition of Pop III stars. However, most literature defines Pop III stars based on criteria in metallicity.  For instance, \cite{Bond1981} defines Pop III stars as having [Fe/H] $< -3$ while \cite{Komiya_ea2015} found that the metallicity of Pop III stars could span a wide range between ${ -8 \lesssim {\rm[Fe/H]} \lesssim -2}$ depending on the merging history of the host halos.  Other authors use $Z < 10^{-3} - 10^{-5} Z_\odot$ where $Z$ is the metal fraction \citep{Bromm_ea2001, Schneider_ea2002, Jaacks_ea2018}. In our work, we shall define a generation of Pop III stars as a class of collapsed stellar objects with low metallicity forming when the host halo met the conditions described in section~\ref{sec:sfrd}.

The organization of this article is as follows; In section~\ref{sec:imf} we give an introduction to the stellar initial mass function (IMF) which is used to calculate the abundance of Pop III stars. In section~\ref{sec:sfrd}, we develop a theoretical formalism to calculate the star formation rate density (SFRD) of Pop III stars, matching our calculations to a simulation result. In section~\ref{sec:evs}, we give an introduction to extreme-value statistics and, in particular, the Generalised Extreme Value approach.  Our main results are given in section~\ref{sec:result} and further implications are discussed in  section~\ref{sec:discussion}.

\section{The Stellar Initial Mass Function}
\label{sec:imf}

The stellar initial mass function (hereafter IMF) is an important  tool in the modelling of stellar abundances. The IMF expresses the number of stars (of a certain type at a fixed time) as a function of their mass. The IMF was first empirically proposed by \cite{Salpeter1955} in the power-law form $\Phi(M) \equiv \D N/\D\log M \propto M^{-\Gamma}$, where $N$ is the number of stars with mass between $\log M$ and $\log M + {\rm d\,log}M$. $\Gamma$ is called the slope. The IMF $\Phi(M)$ describes the stellar mass distribution after their formation.  In this work, we will study two IMFs for Pop III stars. First, the log-normal IMF
\begin{equation}
    \label{eq:LogNormFn}
    \diff{N}{\log M} \propto \exp\bkt{-\bkt{\frac{\log M - \log M\sub{char}}{\sqrt2\sigma}}^2},
\end{equation}
where $M\sub{char}$ is the characteristic mass of Pop III stars, and $\sigma$ is the spread of the mass around $M\sub{char}$. The log-normal IMF was introduced in the pioneering work of \cite{Miller_Scalo1979} who found the form to be a good fit to observation assuming simple models of star birthrates. Our second IMF model is  the \cite{Chabrier2003} IMF
\begin{equation}
    \label{eq:chab}
    \diff{N}{\log M} \propto M^{1-\alpha} \exp\bkts{-\bkt{\frac{M\sub{char}}{M}}^\beta},
\end{equation}
with parameters $\alpha, \beta$ and $M\sub{char}$. This IMF has an interesting flexibility in that it resembles the log-normal IMF for small $M$ and approaches the power-law form $M^{1-\alpha}$ for large $M$ (or when $\beta=0$). An IMF comprising a log-normal body and a power-law tail is expected in a broad class of star-formation scenarios \citep{Basu_Jones2004}. We note that the extreme-value framework that we will present is not limited to these IMFs.

The normalisation of the IMF is usually left unspecified in previous work on stellar population.  Some authors treat $\Phi$ as a probability distribution (so that $\int \D \log M\, \Phi = 1$), and normalise the number count $N$ instead. Alternatively, one can also normalise the IMF using the total stellar mass, meaning that
$
    M_{\rm *}\super{\rm total} = \int_0^{\infty} {\rm d}\log M M \Phi(M).
$
Both normalisation methods depend on the measurement of either the stellar number counts or the total stellar mass for all possible masses of Pop III stars. Since direct observational constraints of Pop III stars are not yet feasible with current technologies, IMF normalisation with these methods are unreliable at best.

In this work, we propose another method of normalising the IMF using SFRD for which we have data from simulations \citep{Gessey-Jones_ea2022} (hereafter GJ22) which applied the star formation model from \cite{Magg_ea2022}. The normalised IMF is necessary for calculating the distribution of the most massive Pop III stars using extreme-value statistics. We discuss the normalisation method in the next section.

\section{Pop III Star Formation Rate Density - a new approach}
\label{sec:sfrd}

We shall develop a methodology to calculate the star formation rate density of Pop III stars based on the modelling of dark matter haloes \citep{Press_Schechter1974} and their cooling temperatures and time-scales \citep{Tegmark_ea1997}.  In our methodology, we propose that the total  density of Pop III stars at redshift $z$ is given by
\begin{equation}
    \label{eq:rhostar3}
    \rho\sub{\rm *, III}(z) = f\sub{\rm *, III} \frac{\Omega\sub{\rm b}}{\Omega\sub{\rm m}} \int_{M\sub{\rm crit}(z)}^{\infty} {\rm d} M M \frac{{\rm d} n}{{\rm d} M}(M, z),
\end{equation}
where $f\sub{\rm *,III}$ is the Pop III star formation efficiency parameter, ${\rm d}n/{\rm d}M$ is the halo mass function, and $n(M,z)$ is the number density of halo mass $M$ at redshift $z$. $\Omega\sub{\rm b}$ and $\Omega\sub{\rm m}$ are respectively the baryonic and total matter density parameters at the present epoch. 

$M\sub{crit}(z)$ is the critical minimum cooling mass of the host halo, given by \citep{Blanchard_ea1992, Tegmark_ea1997}:
\begin{equation}
    \label{eq:mcrit}
    M\sub{crit}(z) = 1.0 \times 10^6 \,{\rm M}_\odot\bkt{\frac{T\sub{crit}}{10^3\,{\rm K}}}^{3/2} \bkt{\frac{1 + z}{10}}^{-3/2}.
\end{equation}
Haloes with mass below $M\sub{crit}(z)$  cannot efficiently dissipate their kinetic energy and become self-gravitating within a Hubble time.  Our assumption is that once $M$ exceeds $M\sub{crit}$, star formation will become effective. The value $T\sub{\rm crit}=2,200\,{\rm K}$ (from considering molecular hydrogen cooling at redshift $z \sim 10$)
will be used \citep{Hummel_ea2012, Magg_ea2022}.

We further assume that $f\sub{\rm *, III}$ is constant  during the epoch where the stellar formation is  dominated by Pop III stars (the effect of time-dependent $f\sub{\rm *, III}$ will be discussed later in section~\ref{sec:discussion}).  The redshift dependence  of $\rho\sub{\rm *, III}(z)$ therefore only comes from $M\sub{crit}(z)$ and ${\rm d}n/{\rm d}M(M, z)$. 

\begin{figure}\centering
    \includegraphics[width=\columnwidth]{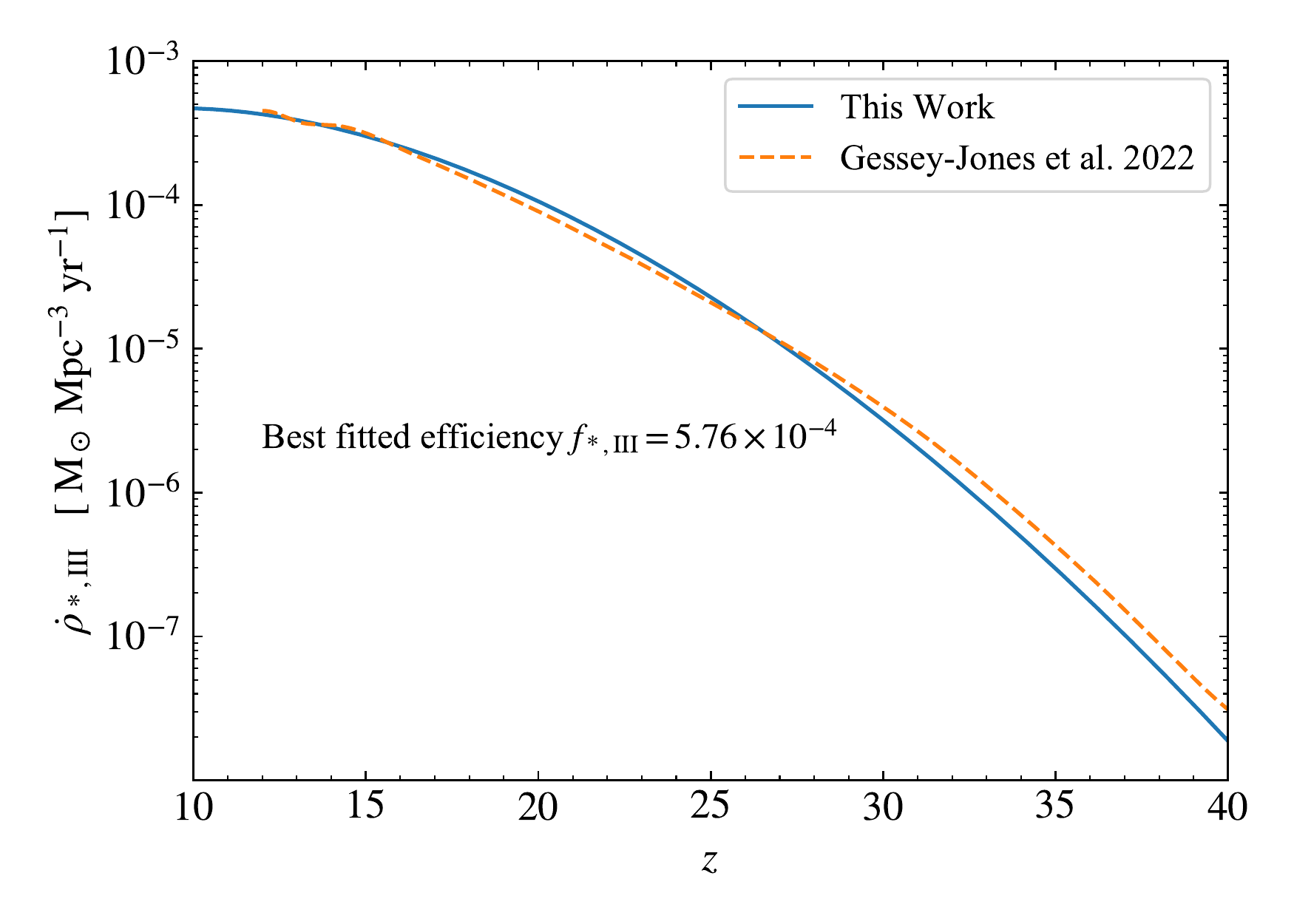}
    \caption{Comparison of the Pop III SFRD.  We compare our Pop III SFRD (solid) to the simulation by \citet{Gessey-Jones_ea2022} (dashed) and find the best-fitting Pop III star formation efficiency $f\sub{\rm *, III} = 5.76\times 10^{-4}$.  The fiducial cosmology is based on Planck 2018 \texttt{Plik} best-fitting parameters. 
    \label{fig:plot_SFRD_GJ}}
\end{figure}

The halo mass function, ${\rm d}n/{\rm d}M$, is defined as the redshift-dependent distribution of the number density of collapsed dark matter haloes per unit mass interval ${\rm d}M$.  It is convenient to express the mass function as
\begin{equation}
    \label{eq:massfunction}
    \diff{n}{M} = \frac{\rho\sub{\rm c}}{M} \diff{\,\ln \sigma^{-1}}{M} f(\sigma),
\end{equation}
where $\rho_{\rm c}$ is the critical density and $\sigma(M, z)$ is the variance of the linear mass density field of mass $M$ at redshift $z$.  The multiplicity function $f(\sigma)$ (also known as the {\it mass fraction} \citep{Jenkins_ea2001}) is defined as the fraction of mass in collapsed haloes per unit interval in $\ln \sigma^{-1}$.  The original Press-Schechter mass fraction, based on spherical collapse, is
\begin{equation}
    \label{eq:massfracps}
    f_{\rm PS}(\sigma) = \sqrt{\frac{2}{\pi}} \frac{\delta_{\rm c}}{\sigma} \exp \bkts{-\frac{\delta_{\rm c}^2}{2\sigma^2}}.
\end{equation}
The Press-Schechter mass fraction tends to underpredict the number of high-mass haloes and overpredict the number of low-mass haloes in the present epoch. It is also notably inaccurate at high redshifts \citep{Lukic_ea2007}.
Here, we will use the Sheth-Tormen mass function based on ellipsoidal collapse \citep{Sheth_Tormen1999}. Its mass fraction is
\begin{equation}
    \label{eq:massfracst}
    f\sub{\rm ST}(\sigma) = A \sqrt{\frac{2a}{\pi}} \frac{\delta\sub{\rm c}}{\sigma} \exp\bkt{\frac{-a \delta\sub{\rm c}^2}{2 \sigma^2}} \bkts{1 + \bkt{\frac{\sigma^2}{a \delta\sub{\rm c}^2}}^p},
\end{equation}
with $A = 0.3222, a = 0.707, \delta\sub{\rm c} = 1.686, p = 0.3$.  This model  gives a good fit to halo abundances in numerical simulations over a wide range of masses and redshifts \citep{Lukic_ea2007}.  Other mass functions have been discussed in the literature, including those by  \citet{Jenkins_ea2001}, \citet{Barkana_Loeb2004}, \citet{Warren_ea2006}, \citet{Reed_ea2007}, \citet{Crocce_ea2010} and \citet{Bhattacharya_ea2011}, with small deviations from the Sheth-Tormen mass function.

We make a simple observation that taking the time derivative of $\rho\sub{\rm *,III}$ (equation~(\ref{eq:rhostar3})) gives the SFRD:
\begin{equation}
    \label{eq:rhostar3dot}
    \dot{\rho}\sub{\rm *, III}(z) = f\sub{\rm *, III} \frac{\Omega\sub{\rm b}}{\Omega\sub{\rm m}} \bkt{\int_{M\sub{\rm crit}(z)}^{\infty} {\rm d} M M \frac{{\rm d} \dot{n}}{{\rm d} M} - \dot{M}\sub{\rm crit} M \frac{{\rm d} n}{{\rm d} M}},
\end{equation}
where a dot denotes time derivative.  The first term on the right involves the time derivative of the mass function in equation~(\ref{eq:massfunction}).  The second term depends on the time derivative of the critical mass in equation~(\ref{eq:mcrit}). 

Our fiducial cosmology is based on Planck 2018 \texttt{Plik} best-fitting parameters \citep{Planck2018_6}.  We compare our calculation with a semi-analytic simulation of GJ22 between $z = 12 - 40$ as shown in Fig.~\ref{fig:plot_SFRD_GJ}. We obtain the best-fitting value $f\sub{\rm *,III} \simeq 5.76 \times 10^{-4},$
which will be important in the extreme-value modelling in the next section. 

\begin{figure}\centering
    \includegraphics[width=\columnwidth]{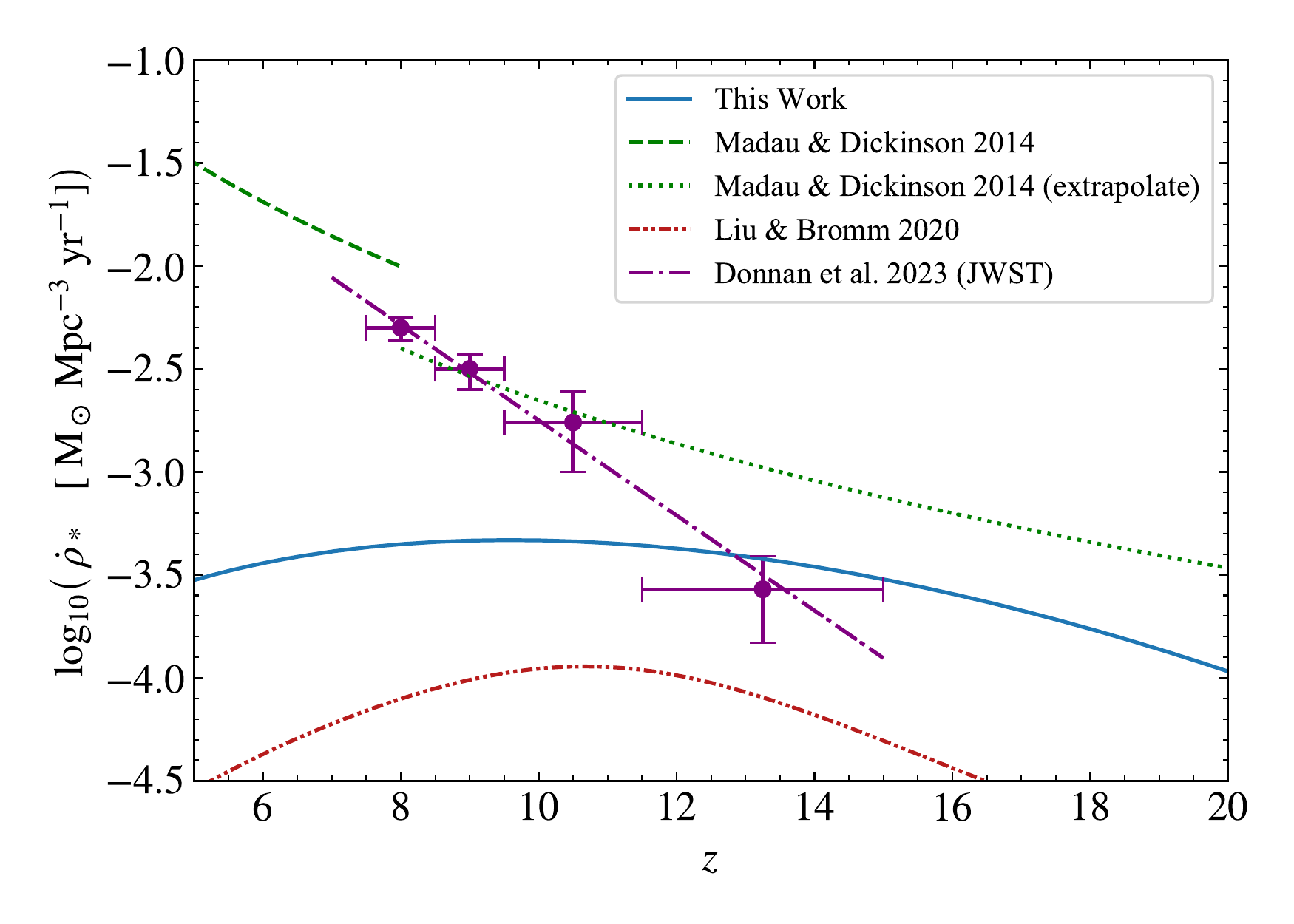}
    \caption{A comparison of SFRD from various models (listed in Table.~\ref{tab:rho_params_fits}) with observation from deep JWST \citep{Donnan_ea2023} (dashdotted). The model of Pop I + II from \citet{Madau_Dickinson2014} (dashed) and its extrapolation with correction factor (dotted) are also shown for comparison (see description in section~\ref{ssec:pop3sfrd}).
    \label{fig:plot_SFRD_Donnan}}
\end{figure}

It is useful to obtain a fitting function of the SFRD. A particular template was suggested by \citet{Madau_Dickinson2014} (hereafter MD14):
\begin{equation}
    \label{eq:madau_fitting}
    \frac{\dot{\rho}_{\rm *}(z)}{{\rm M}_\odot{\rm yr}^{-1} {\rm Mpc}^{-3}} = \frac{a (1 + z)^b}{1 + [(1+z)/c]^d},
\end{equation}
where $a, b, c$ and $d$ are parameters in the fitting function. MD14  proposed this fitting function for Pop I and Pop II SFRD within $z \sim 0 - 8$.  We have also calculated the parameters for the fitting function  and listed them in Table~\ref{tab:rho_params_fits} for our best-fitting $f\sub{\rm *,III}$. The table also compares the values of the fitting parameters and the redshift range of validity from previous authors alongside ours. These models are plotted in Fig.~\ref{fig:plot_SFRD_Donnan}, alongside the observational data from \citet{Donnan_ea2023} (hereafter D23). 

The data are from the \textit{James Webb Space Telescope} and comprise the SFRD from all stellar populations in four redshift bins with mean redshifts $z = 8.0, 9.0, 10.5$ and $13.25$.  Since the \textit{JWST} data include contributions from Pop I and Pop II stars, it is not surprising that our estimate of the Pop III SFRD is below the data in the first three bins where Pop I and Pop II contributions to the SFRD are dominant.  However, our Pop III SFRD agrees with the last bin at mean redshift $z = 13.25$ where the SFRD contribution is dominated by Pop III stars.

\begin{table*}
    \centering
    \caption{Fitting parameters for for the SFRD using the functional form  (\ref{eq:madau_fitting}). The values of the parameters $a, b, c$ and $d$ for SFRD are listed alongside the redshift range and stellar types for which they are valid.}
    \label{tab:rho_params_fits}
    \begin{tabular}{lcccccr}
    \hline\hline
    Reference & Redshift range & Type & \multicolumn{4}{c}{Fitting parameters} \\[0.5em]
    \cline{4-7}
    &&& $a$ & $b$ & $c$ & $d$ \\
    &&& ${\rm M}_\odot$ Mpc$^{-3}$ yr$^{-1}$ & & & \\
    \hline\hline
    This work & 6 -- 20 & Pop III & 250.16 & -4.744 & 14.74 & -5.60 \\
    \citet{Madau_Dickinson2014} & 0 -- 8 & Pop I \& II & 0.015 & 2.7 & 2.9 & 5.6 \\
    \citet{Liu_Bromm2020} & 4 -- 24 & Pop III & 765.7 & -5.92 & 12.83 & -8.55 \\
    \hline\hline
    \end{tabular}
\end{table*}

We can now use our SFRD calculation to normalise the IMF by equating equation~(\ref{eq:rhostar3dot}) to the total number of stellar mass per unit time as determined by the stellar IMF $\Phi(M)$
\begin{equation}
    \label{eqPhi}
    \dot{\rho}\sub{\rm *, III} = A(z) \int_{0}^{\infty} {\rm d}\log M M \Phi(M),
\end{equation}
where $A(z)$ is a redshift-dependent factor that normalises the first moment of the IMF per unit volume per unit time. 

Once $A(z)$ is obtained, the IMF is normalised, and we can write down the number density of Pop III stars above mass $M$ at redshift $z$ (denoted $n(>M, z)$) as 
\begin{equation}
n(>M, z)=A(z)\diff{t}{z}\int_{M}^{\infty} {\rm d}\log M^\prime  \Phi(M^\prime). \label{ngmz}
\end{equation}
We will use this expression to calculate the mass of extreme Pop III stars.

\section{Extreme-Value Statistics}
\label{sec:evs}

Our tool for quantifying the abundances of the most massive Pop III stars is extreme-value statistics. In particular, we will appeal to the generalised extreme-value (GEV) formalism - also known as the block maxima method. The quantity of interest is the probability distribution of block maxima, where a block is a population sample within a fixed volume. After dividing the data into $N$ non-overlapping blocks, we collect the maximum value from each block. Under generic assumptions, the large-$N$ limit (after a certain scaling) is one of three types: the Gumbel, Fr\'echet or Weibull distribution. This result is the celebrated Fisher-Tippett-Gnedenko theorem (analogous to the Central Limit theorem), which plays a key role in  many real-world  applications of extreme-value statistics. For an introduction to the GEV approach in extreme-value statistics, see \cite{deHaan2006, Gomes_Guillou2015}. 

The GEV approach has previously been used to quantify the abundances of the most massive galaxy clusters \citep{Davis_ea2011, Waizmann_ea2012, Chongchitnan_Silk2012} and primordial black holes \citep{Kuhnel_Schwarz2021}. We believe this work is the first time the GEV formalism has been applied to Pop III stars. 

Starting with the number density $n(>M,z)$ in equation~(\ref{ngmz}),  we can calculate the  number density of Pop III stars of mass exceeding $M$ in the entire redshift range $z\in[z_0,z_1]$ as
\begin{equation}
	n(>M)=\int_{z_0}^{z_1}\D z\, n(>M, z).
\end{equation}

Now consider the probability that a region of volume $V$ contains Pop III stars of mass not exceeding $M$. In other words, we are interested in the probability that no Pop III stars of mass $>M$ are found in the volume $V$. In the large volume limit, this probability can be described by the cumulative distribution function (cdf) of the Poisson form  \citep{White1979, Davis_ea2011}
\begin{equation}
	P_0(M)=\exp\bkt{-n(>M) V}.\label{cdf}
\end{equation}

By differentiating this cdf with respect to $M$, we obtain the pdf of the maximum mass Pop III stars within volume $V$. 

In the limit that the Fisher-Tippett-Gnedenko theorem applies, we can equate the cdf (\ref{cdf}) with the GEV distribution 
\begin{equation}
	G(M)=\begin{cases} \exp\bkt{-(1+\gamma y)^{-1/\gamma}}  & (\gamma\neq0),\\
\exp(-e^{-y})& (\gamma =0),\end{cases}
\end{equation}
where $y:=(\log\sub{10}M-\alpha)/\beta$ is the scaled logarithmic mass. The parameter $\gamma$ determines which of the 3 extremal types the block maxima converges to. The Gumbel, Fr\'echet and Weibull distributions correspond to $\gamma=0$, $\gamma>0$ and $\gamma<0$ respectively.

The parameter $\gamma$ as well as the scaling constants $\alpha$ and $\beta$ can be determined as follows. By Taylor expanding the cdf $P\sub{0}(M)$ and the GEV $G(M)$ around the peak $M\sub{peak}$ of the pdf to cubic order, we equate terms and find that  $\alpha, \beta,\gamma$ are given in terms of the redshift-averaged number density $n(>M)$ as:
\begin{align}
    \gamma &= n(>M\sub{peak})V -1\qquad \beta = \frac{(1+\gamma)^{1+\gamma}}{\displaystyle\diff{n}{M}\Big|_{M\sub{peak}} M\sub{peak}V \ln{10} }\notag\\
    \alpha & = \log\sub{10}M\sub{peak}-\frac{\beta}{\gamma}\bkt{(1+
    \gamma)^{-\gamma}-1}.
\end{align}

These values allow us to characterise the extreme-value distribution of Pop III stars. In particular, $\alpha$ corresponds roughly to the peak mass $\log\sub{10} M\sub{peak}$, and $\gamma+1$ is the number count of stars with mass above $M\sub{peak}$. These GEV parameters are important in the modelling of extreme objects because they allow us to venture into domains of small probabilities which would have been numerically prohibitive to calculate otherwise. 

\section{Extreme Pop III stars}
\label{sec:result}

\begin{figure}
    \centering
    \includegraphics[height=2.3in]
    {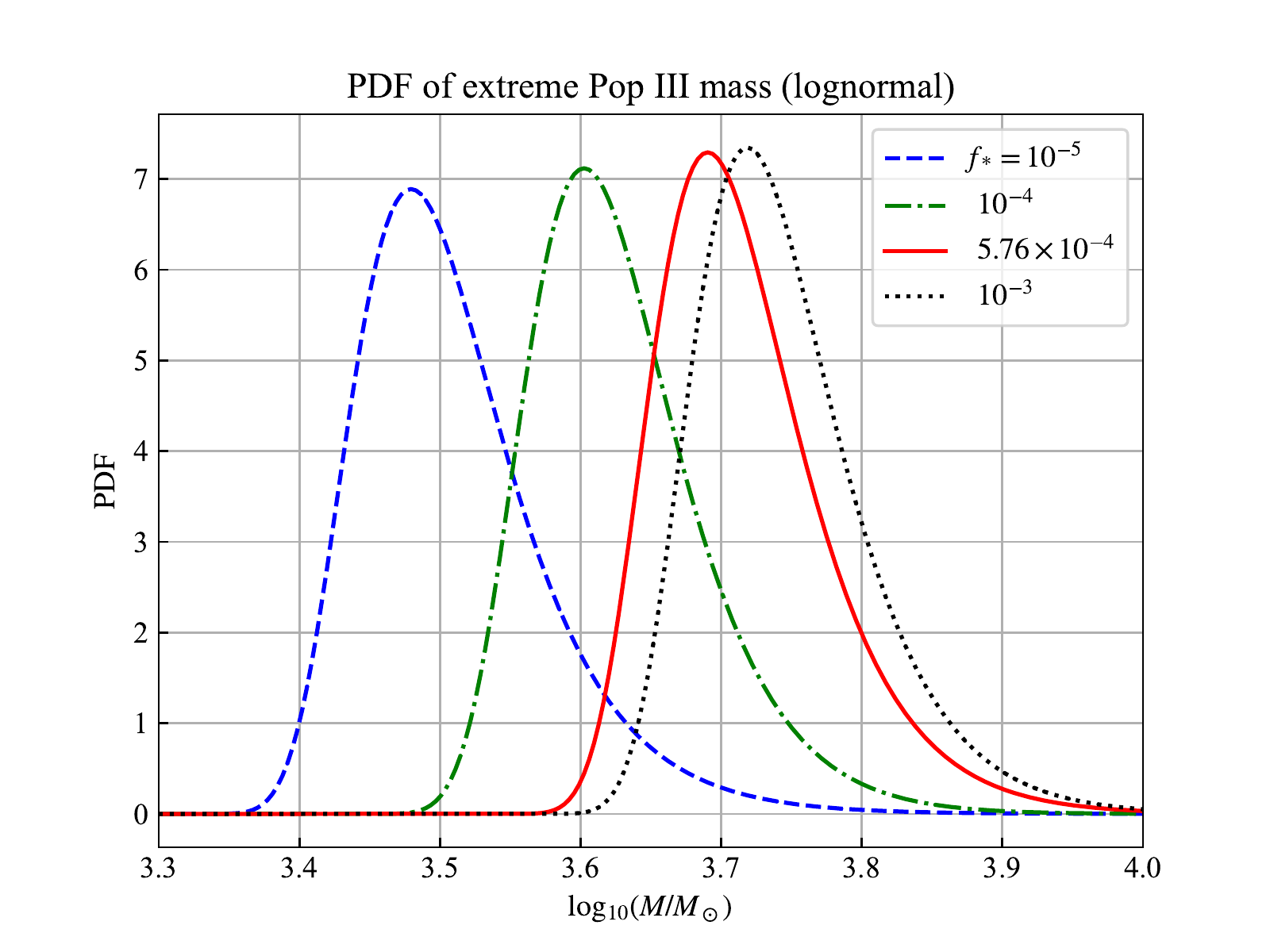}
    \caption{The probability density functions for the extreme-mass Pop III stars for $10< z<20$, assuming the log-normal IMF (eq. \ref{eq:LogNormFn})  with $\sigma^2=1$ and $M\sub{char}=1 \,{\rm M}_\odot$. The 4 curves correspond to  4 values of $f\sub{\rm *, III}$. The curve corresponding to  $f\sub{\rm *, III}=5.76\times10^{-4}$ (solid red line) uses the best-fitting star-formation efficiency obtained from fitting to the simulation of GJ22.} \label{figgumbel2}
\end{figure}

\begin{figure}
    \centering
    \includegraphics[height=2.3in]
    {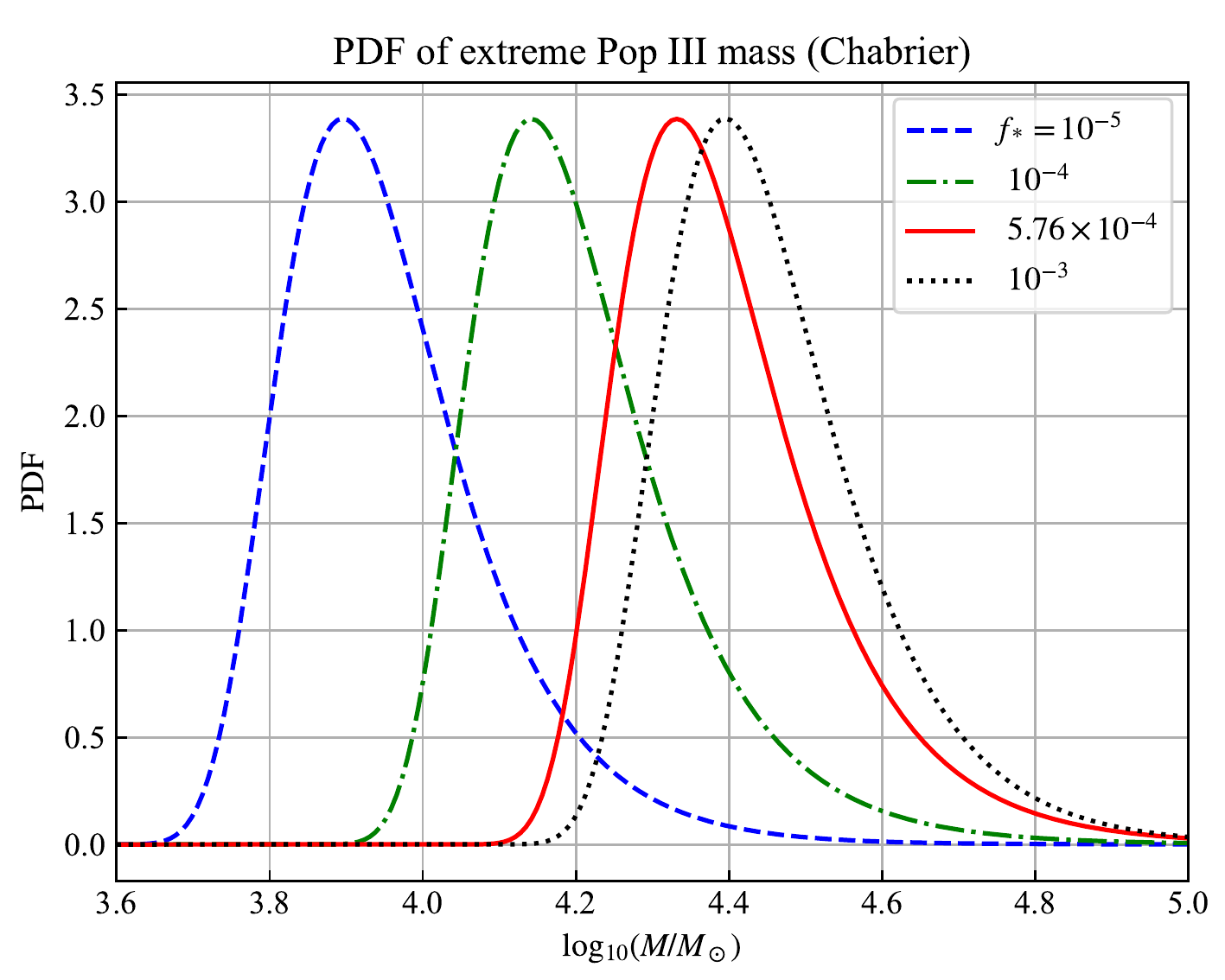}
    \caption{Same as Fig.~\ref{figgumbel2} but for the Chabrier IMF (eq. \ref{eq:chab}) with $\alpha=5$, $\beta=1$ and $M\sub{char}=1 \,{\rm M}_\odot$}. \label{figgumbel2C}
\end{figure}

Two plots of the pdfs for the maximum-mass Pop III stars in the redshift range $z\in[10,20]$ obtained via extreme-value modelling are shown in Fig.~\ref{figgumbel2} and Fig.~\ref{figgumbel2C}. In Fig.~\ref{figgumbel2} we use the log-normal IMF (eq. \ref{eq:LogNormFn}) with parameters $M\sub{char}=1 \,{\rm M}_\odot$ and $\sigma=1$, whilst Fig.~\ref{figgumbel2C} uses the Chabrier model (eq. \ref{eq:chab}) with $\alpha=5, \beta=1$ and $M\sub{char}=1\,{\rm M}_\odot$. In both models, we assume full-sky observation ($f\sub{sky}=1$).  The four curves in each plot correspond to varying values of $f_{\rm *,III}$, with the value $5.76\times10^{-4}$ being the best-fitting value obtained from the SFRD methodology described in section~\ref{sec:sfrd}. For these parameter choices, the pdfs peak at around $\sim 10^{3}- 10^4 \,{\rm M}_\odot$ for the best-fit $f\sub{*,\rm{III}}$. The possibility of such large values of extreme-mass Pop III stars has been hypothesized previously \citep{Haemmerle_ea2018}.

Next, we found that the value of the GEV parameter $\gamma$ (which determines the extremal type) are typically small ($|\gamma|\lesssim0.05$) with the possibility of both positive and negative values. Essentially the pdfs are well described by the Gumbel distribution ($\gamma=0$). This is similar to the conclusion in \cite{Davis_ea2011} who studied the GEV fit for massive clusters. This conclusion is also consistent with the theoretical result that extreme values from the log-normal distribution follow the Gumbel distribution in the limit that the Fisher-Tippett theorem holds. See \cite{Embrechts_ea1997} for details.

In addition, we note that in both models increasing $f\sub{\rm *, III}$ by an order of magnitude shifts the pdfs towards higher extreme masses by roughly a factor of 2. Overall the extreme-mass predictions are quite robust against changes in $f\sub{\rm *, III}$.

\begin{figure}
    \centering
    \includegraphics[height=2.6in]
    {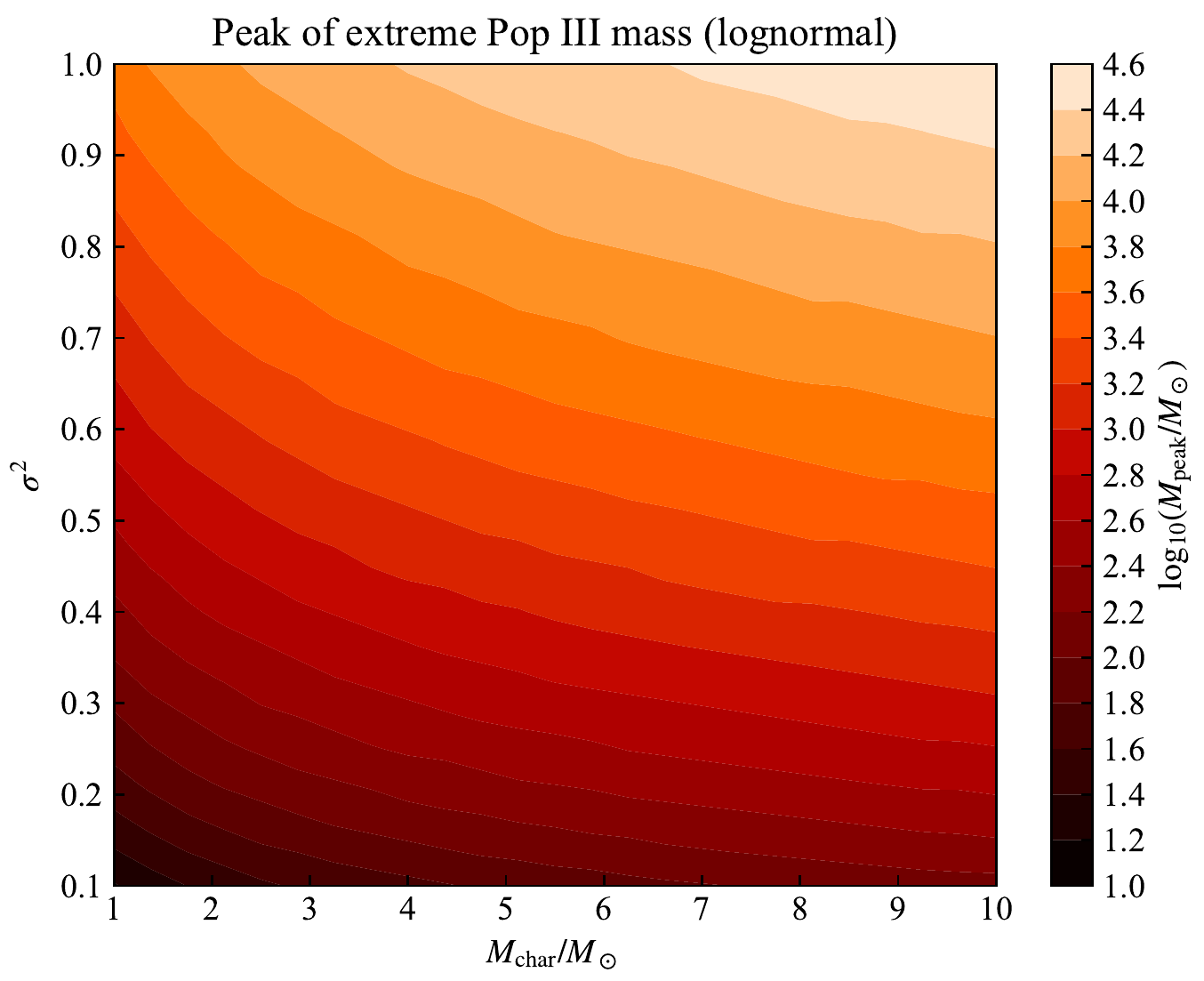}
 \caption{Heat map showing $M\sub{peak}$, the peak of the extreme-value PDF (in $z\in[10,20]$) as a function of the log-normal IMF (eq. \ref{eq:LogNormFn}) with parameters $\sigma^2$  and $M\sub{char}$, whilst fixing $f\sub{\rm *, III}=5.76\times10^{-4}$.} \label{figfull}
\end{figure}

\begin{figure}
    \centering
    \includegraphics[height=2.6in]
    {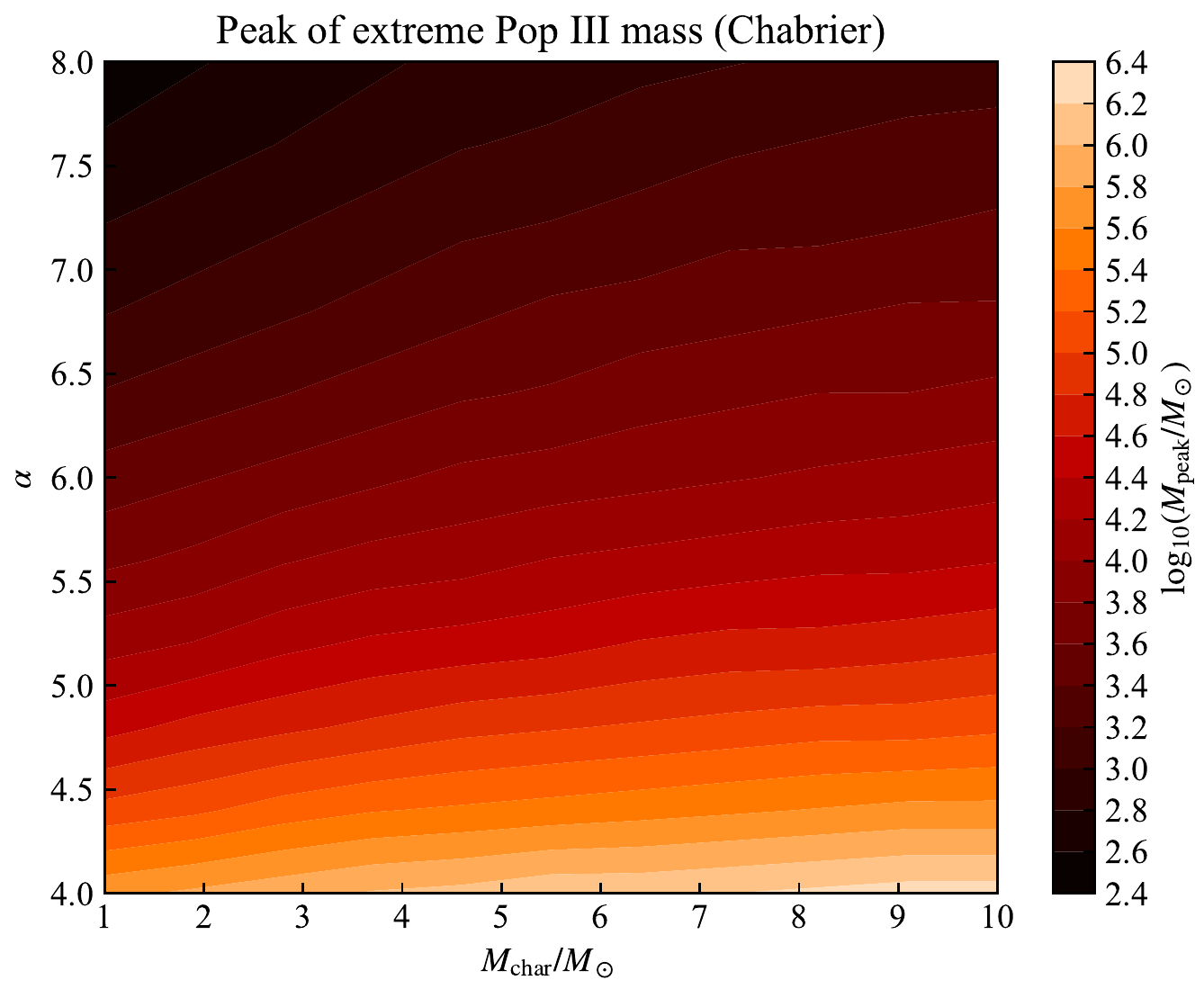}
 \caption{Heat map showing $M\sub{peak}$, the peak of the extreme-value PDF (in $z\in[10,20]$) as a function of the Chabrier IMF (eq. \ref{eq:chab}) with parameters $\alpha$  and $M\sub{char}$, whilst fixing $\beta=1$ and $f\sub{\rm *, III}=5.76\times10^{-4}$.} \label{figfullchab}
\end{figure}

However, the extreme-mass predictions are much more sensitive to changes in some of the parameters of the IMF (and indeed, the form of the IMF itself). We argue that the extreme-value formalism can be used to constrain the model parameters by considering the prediction of the extreme masses of Pop III stars. This is demonstrated in Fig.~\ref{figfull} and Fig.~\ref{figfullchab}  in which we vary the IMF parameters and note the peak of the extreme-value pdf.

In Fig.~\ref{figfull}, we vary $M\sub{char}$ against $\sigma^2$ in the log-normal model. In \ref{figfullchab}, we vary $M\sub{char}$ against $\alpha$ in the Chabrier model (fixing $\beta=1$ - the extreme peaks are insensitive to changes in $\beta$).  In both contour plots, we use the best-fit value of $f\sub{\rm *, III}=5.76\times10^{-4}$. We see that a wide range of extreme Pop III masses $\gtrsim 10^3-10^4 \,{\rm M}_\odot$ are possible. Portions of such IMF parameter space can be effectively ruled out with future observations of massive Pop III stars.

We observe that extreme masses of order $\sim10^3-10^4 \,{\rm M}_\odot$ arise naturally out of the EVS formalism. Calculating the number density $n(>M)$ reveals that Pop III stars of mass $\gtrsim 10^3 \,{\rm M}_\odot$ can in fact form in significant abundances in a wide range of parameter space.  For instance, in the log-normal model with $M\sub{char}=1 \,{\rm M}_\odot$, taking  $\sigma\gtrsim0.7$ yields $n(>10^3 \,{\rm M}_\odot)$ exceeding $10^{-9}\, \text{Mpc}^{-3}$.  This translates to a number count of $N\gtrsim10^3$ objects in $10<z<20$. Such an abundance of massive Pop III stars is ideal for seeding massive black holes at high redshifts, in addition to black holes of primordial origin \citep{Chongchitnan_ea2021}.

\section{Conclusion and discussion }
\label{sec:discussion}

\subsection{Pop III SFRD}
\label{ssec:pop3sfrd}

We have presented a novel methodology for calculating the Pop III SFRD and star formation efficiency $f\sub{\rm *, III}$ (equation~(\ref{eq:rhostar3dot})). We assumed that $f\sub{\rm *, III}$ is constant which is a plausible assumption since the metallicity of Pop III stars varies slowly. \cite{Jaacks_ea2018} has shown that the mean metallicity $Z$ rises smoothly from $z \simeq 25$ reaching $Z\sub{crit} = 10^{-4}Z_\odot$ at $z \simeq 7$ where $Z\sub{crit}$ is the transition metallicity for the Pop III to Pop II stars. Thus, we would expect $f\sub{\rm *, III}$ to also increase slowly with time.  To implement this, we could add an extra term involving $\dot{f}\sub{\rm *, III}$ in equation~(\ref{eq:rhostar3dot}), where $\dot{f}\sub{\rm *, III}$ is small. Even with a varying efficiency, we  expect the effect on the SFRD to be small.

The assumptions made in our methodology are sufficient for a good fit to be achieved in comparison with the simulation from GJ22 in Fig.~\ref{fig:plot_SFRD_GJ} and observational data from deep JWST (D23) in Fig.~\ref{fig:plot_SFRD_Donnan}.

We also gave a fitting function for our SFRD in equation~(\ref{eq:madau_fitting}), shown in Fig.~\ref{fig:plot_SFRD_Donnan} along with those of previous authors, including MD14 and \citet{Liu_Bromm2020}. It is interesting to note that MD14 proposed this fitting function for Pop I and Pop II stars, and therefore the function has a limited validity range $z = 0 - 8$.  A direct extrapolation of MD14 to higher redshifts overestimates the SFRD; however, as anticipated by \citet{Shapley_ea2023}, the conversion factor between H$\alpha$ luminosity and SFRD should be lower by a factor of $\sim 2.5$.  We apply the correction factor of 2.5 to MD14 extrapolation and obtained an improved consistency with the JWST data.

We provide our fitting function for the Pop III SFRD in Table~\ref{tab:rho_params_fits}. It only matches our Pop III SFRD well within $z = 6 - 20$, beyond which  the functional form fails to capture the rapid decrease in the SFRD at higher redshifts. 
 Nevertheless, our fitting function should be useful for  calculations involving the total SFRD. 
 
 In comparison, D23 has also provided a simple fitting function for their data (in Fig.~\ref{fig:plot_SFRD_Donnan}) with limited validity range as
\begin{equation}
    \label{eq:donnan_fit}
    \log\sub{10} \dot{\rho}\sub{*} = (-0.231 \pm 0.037) \times z - (0.43 \pm 0.3),
\end{equation}
where $\dot{\rho}\sub{*}$ is the SFRD in unit of ${\rm M}_\odot\, {\rm Mpc}^{-3}\, {\rm yr}^{-1}$.  The validity range of the fitting function in equation~(\ref{eq:donnan_fit}) is only from $z \sim 7 - 13$.  We recommend using the fitting function in equation~(\ref{eq:donnan_fit}) within its validity range together with our fitting function at higher redshifts for the total SFRD (See Table~\ref{tab:rho_params_fits}). 

\subsection{Extreme Pop III stars}
\label{ssec:massivepop3}

We applied the SFRD methodology to the calculation of the probability distribution of the most massive Pop III stars expected in the redshift range 10 to 20. Adoption of a functional form of the stellar IMF allowed the IMF to be normalised, and the number density of Pop III stars can then be calculated.  The extreme-value pdf were then derived as shown in Fig.~\ref{figgumbel2} and \ref{figgumbel2C}. We demonstrated that for a wide range of parameter values in the log-normal and Chabrier IMF, extreme Pop III stars of mass $\sim10^3 -10^4\, {\rm M}_\odot$ arose naturally, and even higher masses are achievable. We conclude that extreme-value statistics can help effectively constrain the IMF of Pop III stars. In addition, Extreme Pop III stars are a viable channel for producing high-redshift quasars and massive black-holes whose gravitational-wave signals  may be detectable by \textit{LIGO}\footnote{\url{https://www.ligo.org}} or the next generation of gravitational wave observatories such as the Einstein Telescope\footnote{\url{https://www.et-gw.eu}} and \textit{LISA}\footnote{\url{https://lisa.nasa.gov}}.  In short, our predicted extreme Pop III stars are plausible candidates for seeding SMBH at high redshifts. They form $10^3-10^4 \,{\rm M}_\odot$ BH at early epochs, allow the required numbers of e-folds of growth by Eddington-limited accretion, and are rare but still sufficiently numerous to solve the seeding problem for high-redshift quasars.

\section*{Acknowledgements}

We would like to thank Thomas Gessey-Jones for his kind provision of the data and useful suggestions.  Without the data, our work would be far from complete.  We also thank  Suraphong Yuma and Tirawut Worrakitpoonpon for their suggestions.  ``This work was supported by Naresuan University (NU), and the National Science, Research and Innovation Fund (NSRF). Grant NO. R2566B091.

\section*{Data Availability}

The data underlying this article will be shared on reasonable request to the corresponding author.

\bibliographystyle{mnras}

\bsp
\label{lastpage}
\end{document}